\documentclass[fleqn,usenatbib]{mnras}

\usepackage{newtxtext,newtxmath}
\usepackage[T1]{fontenc}

\DeclareRobustCommand{\VAN}[3]{#2}
\let\VANthebibliography\thebibliography
\def\thebibliography{\DeclareRobustCommand{\VAN}[3]{##3}\VANthebibliography}

\usepackage{graphicx}	% Including figure files
\usepackage{amsmath}	% Advanced maths commands
\usepackage{gensymb}
\usepackage{multirow}
\usepackage{xspace}

\usepackage[normalem]{ulem}

\title{Gamma-ray flares from pulsar wind nebulae in the Large Magellanic Cloud}
\author[Nizamov \& Pshirkov]{
B. A. Nizamov,$^{1}$\thanks{E-mail: nizamov@physics.msu.ru (BAN)}
M. S. Pshirkov$^{1,2,3}$
\\
$^{1}$
Sternberg Astronomical Institute, Moscow State University, 119992, 
Universitetskiy Prospekt, 13, Moscow, Russia\\
$^{2}$
Institute for Nuclear Research of the Russian Academy of Sciences, 117312, Moscow, Russia\\
$^{3}$
P. N. Lebedev Physical Institute of the Russian Academy of Sciences, Pushchino Radio Astronomy Observatory,
Pushchino 142290, Russia\\
}

\date{Accepted XXX. Received YYY; in original form ZZZ}

\pubyear{2021}

\begin{document}
\label{firstpage}
\pagerange{\pageref{firstpage}--\pageref{lastpage}}
\maketitle

\begin{abstract}
High-energy radiation of young pulsar wind nebulae (PWNe) is known to be variable. This is exemplified by the Crab nebula
which can undergo both rapid brightenings and dimmings. Two pulsars in the Large Magellanic Cloud, PSR~J0540--6919 and
PSR~J0537--6910 are evolutionally close to Crab, so one may expect the same kind of variability from the PWNe around them. In
this work we search for flaring activity in these PWNe in gamma rays using the data from the Fermi Large Area Telescope in
the range 100~MeV--10~GeV collected from August 2008 to December 2021. We construct light curves of these sources in the
three bands, 100--300~MeV, 300--1000~MeV and 1--10~GeV with one week resolution. We find evidence of flaring activity in
all the bands, in contrast with Crab, where no flares at $E$>1 GeV were observed. Analysis of the flaring episode in the
100--300 and 300--1000~MeV bands indicates that the flux of one of the PWNe could grow by a factor of $\approx 5-10$ and
the statistical significance of the flare detection reaches 6$\sigma$. We are not confident about which of the two PWNe
flared because of their proximity in the sky. However, in the 1--10~GeV band where the angular resolution of LAT is better,
we find several episodes of enhanced brightness in both the PWNe. We check possible contaminants which could be responsible
for the observed variability, but find their contribution not to be relevant.
\end{abstract}


%%%%%%%%%%%%%%%%%%%%%%%%%%%%%%%%%%%%%%%%%%%%%%%%%%

%%%%%%%%%%%%%%%%% BODY OF PAPER %%%%%%%%%%%%%%%%%%

%%%%%%%%%%%%%%%%%%%%%%%%%%%%%%%%%%%%%%
%%%%%%%%%%% INTRODUCTION %%%%%%%%%%%%%
%%%%%%%%%%%%%%%%%%%%%%%%%%%%%%%%%%%%%%

\section{Introduction}
Pulsar wind nebulae (PWNe) form inside supernovae remnants when the flow of relativistic
leptons  from the pulsar expands into the material of the explosion. Their radiation is produced by the 
synchrotron and inverse Compton  emission of these particles re-accelerated at the termination shock -- the boundary between the freely expanding
wind and the surrounding supernova ejecta \citep{Gaensler2006}.
A sub-population of the youngest PWNe with ages of the order of thousands of years is of great interest.
Pulsars generating them have the highest spin-down power, so naturally these PWNe are often the most luminous ones.
Some of young PWNe are reported to be variable. Morphological  changes in  Crab PWN torus and jets
were observed in radio, optical and X-ray bands \citep{Hester2002, Bietenholz2004, Weisskopf2011}.
Shifts and changes in brightness in Vela PWN on the time scale of several
months in X-rays were reported in \citep{Pavlov2001}. X-ray brightness of the youngest known PWN, the one inside the supernova remnant Kes~75,
 decreases by tens per cent over a time of several years \citep{Reynolds2018, Ng2008}. It was also shown that the PWN around a young pulsar PSR~J1809–1917 was variable in X rays on a time scale of several months \citep{Klingler2018}. The most remarkable example of PWN variability is probably the famous Crab nebula which demonstrates gradual dimming in X-rays \cite{Wilson-Hodge2011}, strong flares \citep{Buehler2012} and dimmings \citep{Pshirkov2020} in gamma-rays.

Properties of two pulsars in the Large Magellanic Cloud, PSR~J0537--6910 and PSR~J0540--6919 are very close to those of Crab. The ages of Crab, PSR~J0540--6919 and PSR~J0537--6910 are 970, 1140 and 5000~yr, periods are 30, 50 and 16~ms and spin-down powers are $4.5 \times 10^{38}$, $1.5 \times 10^{38}$ and $4.9 \times 10^{38}$~erg/s respectively \citep{Ackermann2015}. 
These objects  could be called  "Crab-like pulsars"  because
of their similar spin-down powers (which are the highest three among the known pulsars) and
other similarities between Crab and especially the "Crab twin" PSR~J0540--6919: the ratio of X-ray to
gamma-ray luminosity is of the order 1; pulse peaks in different energy bands are in phase; the pulsars emit giant
radio pulses and their position  coincides in phase with the peak of high-energy radiation \citep{Takata2017}.

PSR~J0540--6919 is also remarkable for its spin-down rate irregularities \citep{Marshall2015, Ge2019}: in December 2011,
its rotation frequency first derivative in  two
weeks increased in absolute value by 36 per cent. Interestingly, \cite{Ge2019} found that the X-ray brightness of the associated PWN
increased by 32 per cent 143 days after this event.

It is reasonable to assume that the properties of PWNe are linked to the properties of the
pulsars inside them. Therefore, one might expect that the high-energy radiation of PWNe
around PSR~J0537--6910 and PSR~J0540--6919 could be variable like that of the Crab nebula
and, possibly, demonstrate flares. In this paper, we attempt to check this hypothesis.
In Section~\ref{sec:observations} we describe the observations we analyzed, we present the results in
Section~\ref{sec:results} and discuss them in Section~\ref{sec:discussion}. We make conclusions in Section~\ref{sec:conclusions}.

%This is a simple template for authors to write new MNRAS papers.
%See \texttt{mnras\_sample.tex} for a more complex example, and \texttt{mnras\_guide.tex}
%for a full user guide.

%All papers should start with an Introduction section, which sets the work
%in context, cites relevant earlier studies in the field by \citet{Fournier1901},
%and describes the problem the authors aim to solve \citep[e.g.][]{vanDijk1902}.
%Multiple citations can be joined in a simple way like \citet{deLaguarde1903, delaGuarde1904}.

%%%%%%%%%%%%%%%%%%%%%%%%%%%%%%%%%%%%%%
%%%%%%%%%%% OBSERVATIONS %%%%%%%%%%%%%
%%%%%%%%%%%%%%%%%%%%%%%%%%%%%%%%%%%%%%
\section{Observations}\label{sec:observations}
In our analysis we use the data from the Fermi Large Area Telescope collected from August 2008
to December 2021.

We have selected events
that belong to the ”SOURCE” class. The Pass 8R3 reconstruction and \emph{fermitools}\footnote{https://github.com/fermi-lat/Fermitools-conda/} were used. A usual event quality cut, namely that the zenith angle should be less than 90$^{\circ}$
was imposed.
We took a circle of $15\degree$ radius around the position of PSR~J0540--6919  ($\alpha_{J2000} = 85.0887, \delta_{J2000} = -69.3383$) as
our region of interest (RoI). To construct weekly light curves of PSR~J0537--6910 and PSR~J0540--6919, we carry out the standard binned likelihood analysis with \textit{gtlike}.

In the source model, we include all the sources from the 4FGL-DR2 catalogue \citep{4FGL, 4FGLDR2}
within $20\degree$ from PSR~J0540--6919, as well as the Galactic and isotropic backgrounds.
For PSR~J0537--6910 and PSR~J0540--6919  all parameters are set free. For the rest of the sources,
we fit the normalization if the test statistic of the source is greater than 25 which corresponds to $5\sigma$ detection. Otherwise, the spectral parameters are kept fixed. Typically, only few point sources, if any, were fit together with the pulsars
in study. We perform the likelihood analysis in the time bins
of 7 days. In 4FGL, the spectral shape of PSR~J0537--6910 is described by a simple power law, while
that of PSR~J0540--6919 is described by a power law with an exponential cutoff (PLSuperExpCutoff2).
However, if the pulsar produces flares they can have a spectral shape different from that in the catalog,
which is the case in Crab PWN, therefore we decided to construct the light curves in three spectral bands, namely,
100 -- 300~MeV, 300 -- 1000~MeV and 1 -- 10~GeV. Within each of them we approximate the shape of
PSR~J0540--6919 and PSR~J0537--6910 with a simple power law.

When analyzing the light curves of PSR~J0540--6919 and PSR~J0537--6910, we found that in the 100--300 and 300--1000~MeV bands,
one can barely distinguish the two objects. The reason is two-fold: first, the angular distance
between them is only 0.25\degree, while the 68 per cent containment angle of the LAT for
front+back events at 0.1 and 1~GeV is about 5 and 1\degree{} respectively; second, the spectral slopes of PSR~J0540--6919 and
PSR~J0537--6910  are both close to 2 in these bands, so that the spectra are almost identical.
This confusion results in that the photons from one source can be occasionally assigned to the other
leading to a spurious variability in both.
Therefore, we decided to replace both sources with the composite source
located in the middle between them. We compared the light curves of separate sources and that of the
composite  source and found that the sum of the former is close to the latter. It means that the photons
from the pulsars and their nebulae are accounted for in our composite source and are not lost, i.e. wrongly ascribed to other sources.

In the 1--10~GeV band, the two sources can be separated quite well because of better angular resolution of the LAT. 

\section{Results} \label{sec:results}
\begin{figure*}
	\includegraphics{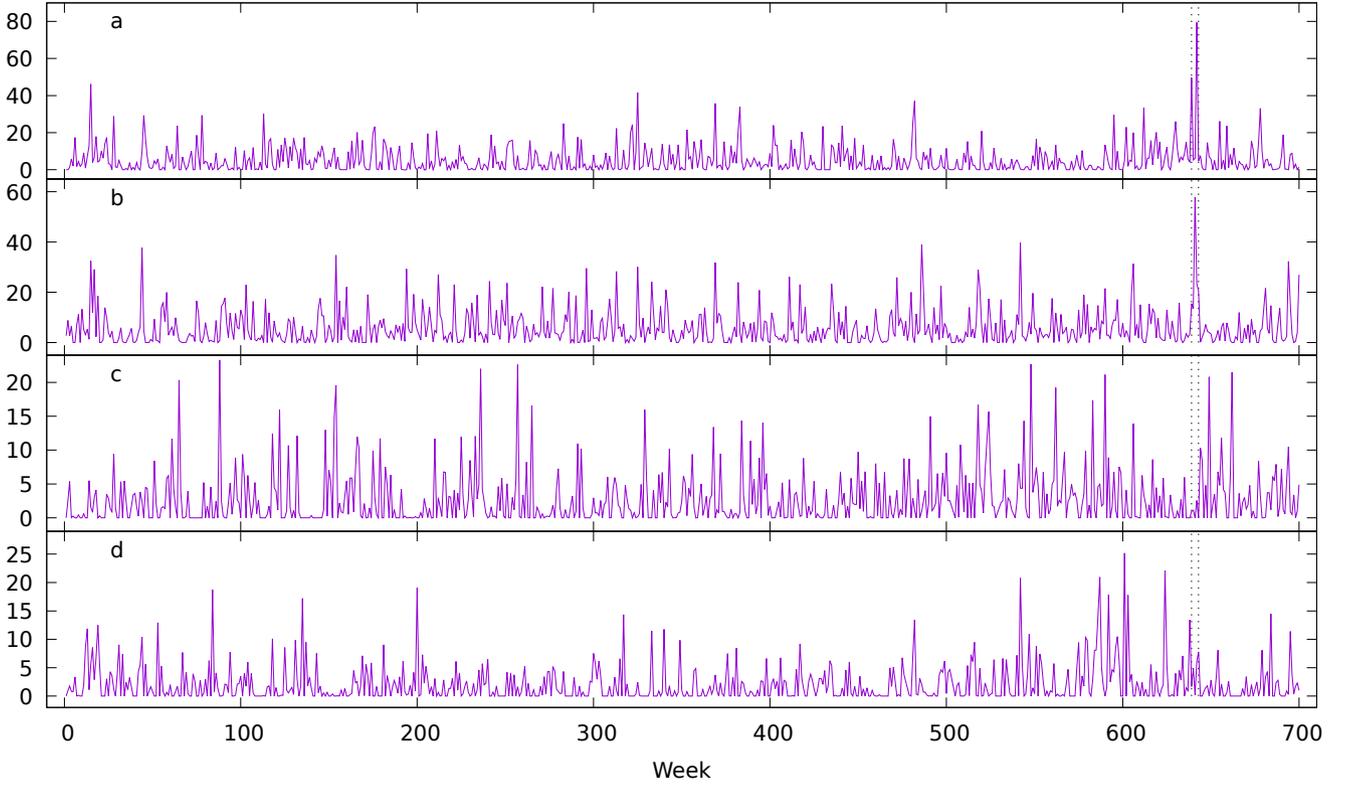}
    \caption{Plots of TS of the composite source in the bands 100--300~MeV (a) and 300--1000~MeV (b); PSR~J0540--6919 (c) and PSR~J0537--6910 (d) in the band 1--10~GeV.
    Vertical dashed lines mark the weeks 639--642.}
    \label{fig:ts_week}
\end{figure*}

In Fig.~\ref{fig:ts_week} we show the test statistic (TS) of the composite source in the bands 100--300 and 300--1000~MeV. In the 1--10~GeV band, the spacial resolution of LAT is sufficient to resolve the two pulsars, and we plot the individual light curves. A remarkable feature is the spike during the weeks 639--642 seen in
the two upper panels of the figure. It probably indicates a flare in one of the sources. We will investigate this
event in more detail in what follows.

\begin{figure*}
	\includegraphics{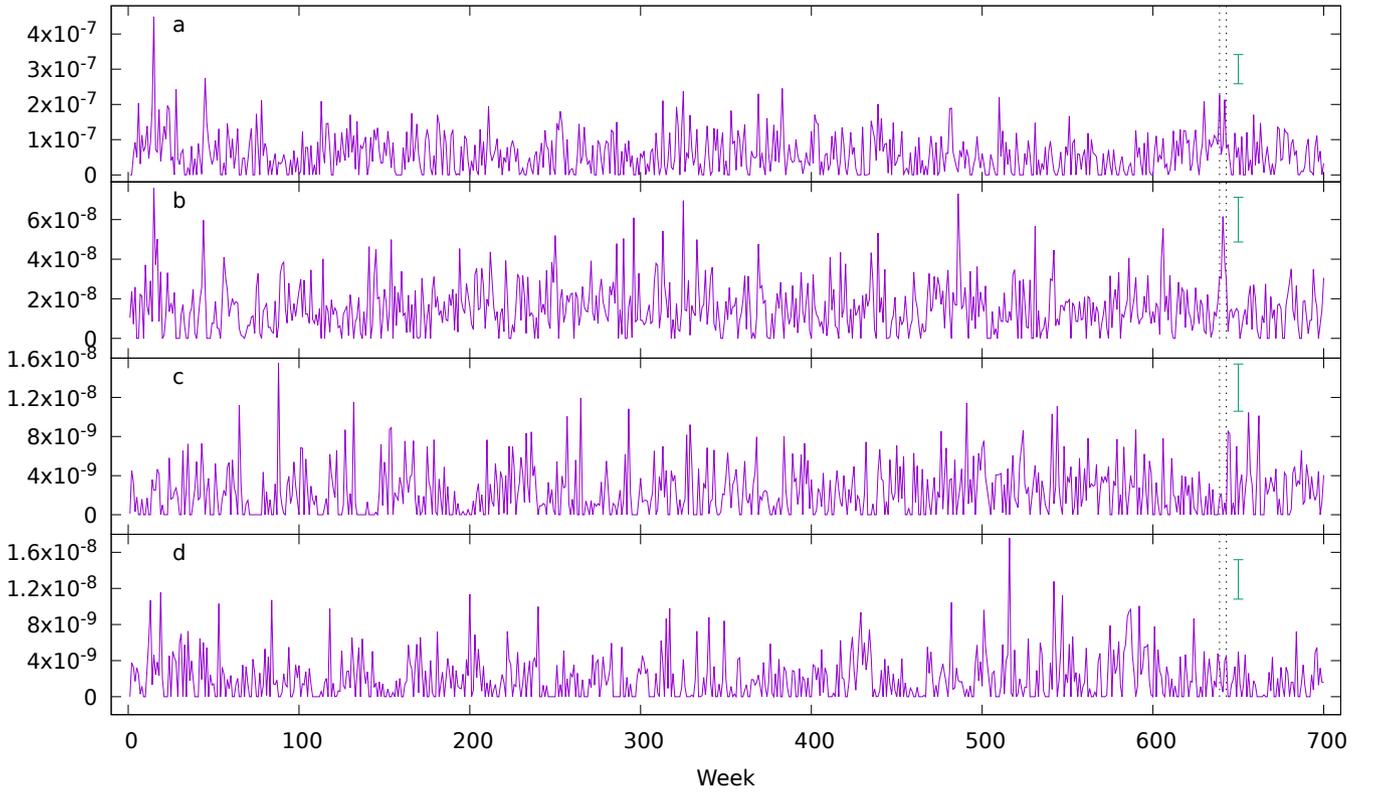}
    \caption{Light curves of the composite source in the bands 100--300~MeV (a) and 300--1000~MeV (b); PSR~J0540--6919 (c) and PSR~J0537--6910 (d) in the band 1--10~GeV.
    The units are photons cm$^{-2}$ s$^{-1}$. The error bars are the median flux errors in the corresponding bands. Vertical dashed lines mark the weeks 639--642.}
    \label{fig:flux_week}
\end{figure*}

Figure~\ref{fig:flux_week} shows the light curves of the sources. For the weekly time bins, the fluxes are determined
with rather large uncertainties, therefore, we do not show the error bars which would obscure the picture. The median flux errors are respectively $4.16\times 10^{-8}$, $1.13\times 10^{-8}$, $2.4 \times 10^{-9}$ and $2.2 \times 10^{-9}$~cm$^{-2}$s$^{-1}$ and shown in the figure. The same burst
is visible again in the upper panels, especially in the panel (b). In the light curves this burst does not stand out so strongly as in the TS plots, but it is the only flare which duration exceeds two weeks. For the sake of robustness, we add this requirement in order to suppress probability of spurious detection from one-week long fluctuations.

Quite a strong flare is seen on the week 15 in the bands 100--300 and 300--1000~MeV. This flare does not originate from the
LMC, and we will return to this point in Section~\ref{sec:discussion}. 
To check the spike in the weeks 639--642 more thoroughly, we developed the following procedure. We select the
photons in 300--1000~MeV range in the time interval of one year centered on the weeks 639-642, but excluding them.
For this interval, we perform the likelihood analysis to obtain the spectral parameters of the brightest
sources.\footnote{For this task, we used the models for PSR~J0537--6910 and PSR~J0540--6919 separately due to the sufficient
photon statistic.} Next, we fix these parameters and use them for the likelihood analysis within the time interval
under investigation, i.e. weeks 639--642. However, along with PSR~J0540--6919 and PSR~J0537--6910 we introduce an additional source to the model. This source represents the
flare, its coordinates coincide with that of PSR~J0540--6919, and its spectrum is described by a power law model. Exact location of the flare  was studied in greater details as well, as described in Section~\ref{sec:discussion}. Thus, the model contains both pulsars with the fixed parameters and the flare source with the parameters set free.\footnote{One could argue that on a four week interval, such a combination is redundant and one composite source would suffice. However, we isolate the \textit{excessive} flux over the quiescent level to show its significance in the next Section.}
The test statistic of this source is found to be 35, corresponding to $\sim6\sigma$ detection, and its flux is $(3.0 \pm 0.6) \times 10^{-8}\text{cm}^{-2}\text{s}^{-1}$.
For comparison, the fluxes of PSR~J0540--6919 and PSR~J0537--6910 within the year are
$(1.24 \pm 0.27) \times 10^{-8}$ and
$(2.7 \pm 2.5) \times 10^{-9}\text{cm}^{-2}\text{s}^{-1}$
respectively, so there is a threefold increase in flux. We did the same calculations in the 100--300~MeV band. Here, the flux of the flare is $(7.78 \pm 2.10) \times 10^{-8} \text{cm}^{-2}\text{s}^{-1}$, while the fluxes of PSR~J0540--6919 and PSR~J0537--6910 are $7.97 \times 10^{-8}$ and $9.7 \times 10^{-9} \text{cm}^{-2}\text{s}^{-1}$ respectively, so the brightening is by factor 2.

The analysis described above has an advantage of improving the photon statistic significantly. While the expected number of photons from both pulsars in the 300-1000~MeV band is about 10 per week, in the four week interval the predicted count number ('npred' parameter) is 44. At the same time, the predicted count number from the flare source is 87. In other words, we observe 131 photons from the pulsars instead of expected 44, the p-value being $\sim 10^{-26}$.

%%%%%%%%%%%%%%%%%%%%%%%%%%%%%%%%%%%%%%%%%
%%%%%%%%%%%%%%% DISCUSSION %%%%%%%%%%%%%%
%%%%%%%%%%%%%%%%%%%%%%%%%%%%%%%%%%%%%%%%%
\section{Discussion} \label{sec:discussion}
One has to
make sure that the variability of the  source under investigation is not caused by  some background source. In the gamma-ray
band, such confusion may arise due to some (maybe unknown) blazar. %\bulat{or a gamma-ray burst?}.
According to the
4FGL-DR2 catalog, the closest blazar to the pulsars in study is an FSRQ PKS~0601--70 which lies 
$2.2\degree$ from PSR~J0540--6919. This blazar is variable: in February 2020, it underwent a major flare \citep{2020ATel13506}.
In the time bins corresponding to this flare (weeks 602--604) we found enhanced flux for PKS~0601--70.  No corresponding enhancement was detected for pulsars.

\begin{figure}
	\includegraphics[width=\columnwidth]{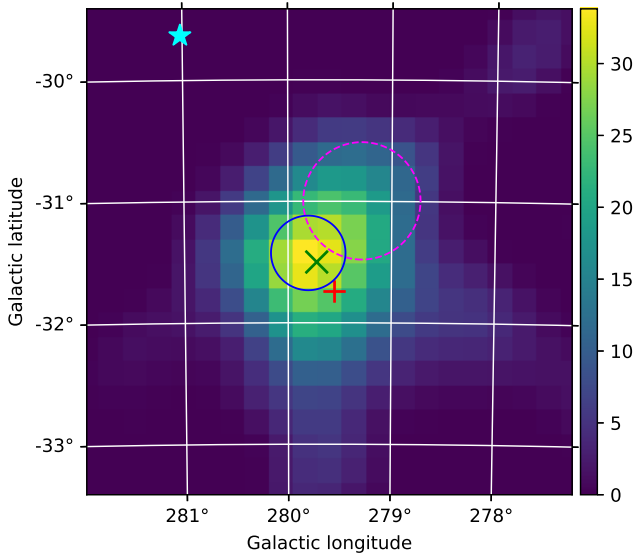}
    \caption{Results of the search for a flaring source in the 300--1000~MeV band. In background is shown the TS map of the
    flare on weeks 639--642. The green cross, red plus sign and cyan star represent PSR~J0540--6919, PSR~J0537--6910 and PKS~0601-70 respectively.
    The magenta dash-lined circle of $29\arcmin$ radius represents the flare on week 15 first reported by \protect\cite{Fermi-flare}. The blue solid circle of $18\arcmin$ radius
    represents the position of the flaring source. The circles correspond to the 95 per cent confidence level.}
    \label{fig:findsrc300}
\end{figure}

Another possible contaminant is an unknown source reported by \cite{Fermi-flare}. In this paper, the first eleven
months of observations of the LMC with the Fermi-LAT were presented. The analysis of variability was also performed, and the authors found a
flare in the fourth month of the observations. It is located at a distance of 47\arcmin{} from PSR~J0540--6919 with
the 95 per cent containment radius of 29\arcmin. The source of the flare could not be identified.
In our analysis, we see this flare on the 15th week in the light curves of the composite source both in 100--300
and 300--1000~MeV bands. Given the absence of any corresponding source in the model and the proximity of the flare to
the pulsars, it is quite natural that \textit{gtlike} assigned the photons of the flare to the composite source.
In principle, this could happen again resulting in a spurious flare detection in the PWNe. To eliminate this
possibility, we performed a search for an unknown source using
\textit{fermipy} \citep{Wood2017}. This is done in two steps. First, a source is located by
constructing a TS map and finding a local maximum in it. Then its position is
refined by scanning the likelihood surface near the initial guess. Once again,
the search is performed in the flare weeks 639--642 with the parameters of
PSR~J0540--6919 and PSR~J0537--6910 fixed at their values in the year around the flare.
The results of the search are presented in Fig.~\ref{fig:findsrc300}.
The green cross and red plus sign indicate the locations of PSR~J0540--6919
and PSR~J0537--6910 respectively; the coordinates and the radius (29\arcmin) of the magenta circle correspond to the
coordinates and their uncertainty of the flare from \citep{Fermi-flare}. The blue circle and its radius (18\arcmin)
represent the position and uncertainty of the flaring source on the weeks 639--642 localized by \textit{fermipy}.
One can see that the position of the flare consistent with the position of PSR~J0540--6919 is preferred over
the position of the unknown flare on the week 15.

\begin{figure}
	\includegraphics[width=\columnwidth]{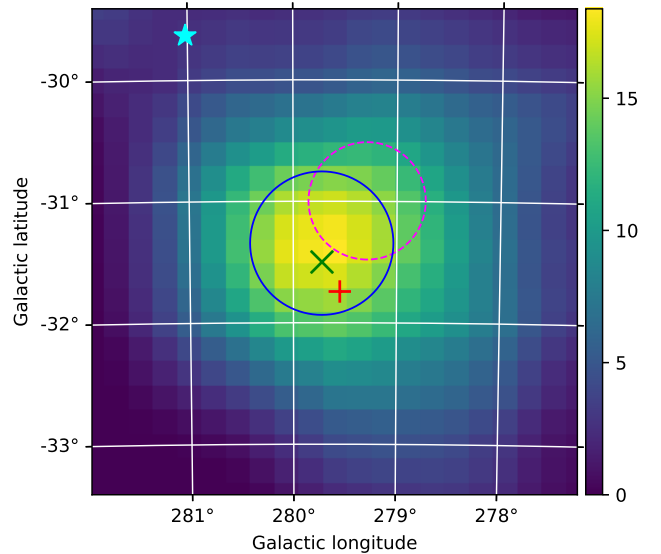}
    \caption{Results of the search for a flaring source in the 100--300~MeV band. The symbols are the same as in Fig.~\ref{fig:findsrc300}. The radius of the blue circle is $35\arcmin$.}
    \label{fig:findsrc100}
\end{figure}

We carried out an analogous analysis also in the bands 100--300~MeV and 1--10~GeV. In the first one, the results are qualitatively identical, but the flaring source is less pronounced and localized worse. The results are shown in Fig.~\ref{fig:findsrc100}.
In the 1--10~GeV band, the analysis could not find any flaring source.

The pulsars are located near a bright extended source, the 30~Dor nebula, and its contribution could affect the photon statistic. However, in the four week interval, the predicted count number from 30~Dor is 58, and these photons come from an area approximately $1\degree \times 1\degree$. It seems unlikely to observe a fluctuation of $\approx 90$ additional photons that would imitate  photons from a point-like source.

\begin{figure}
    \includegraphics[width=\columnwidth]{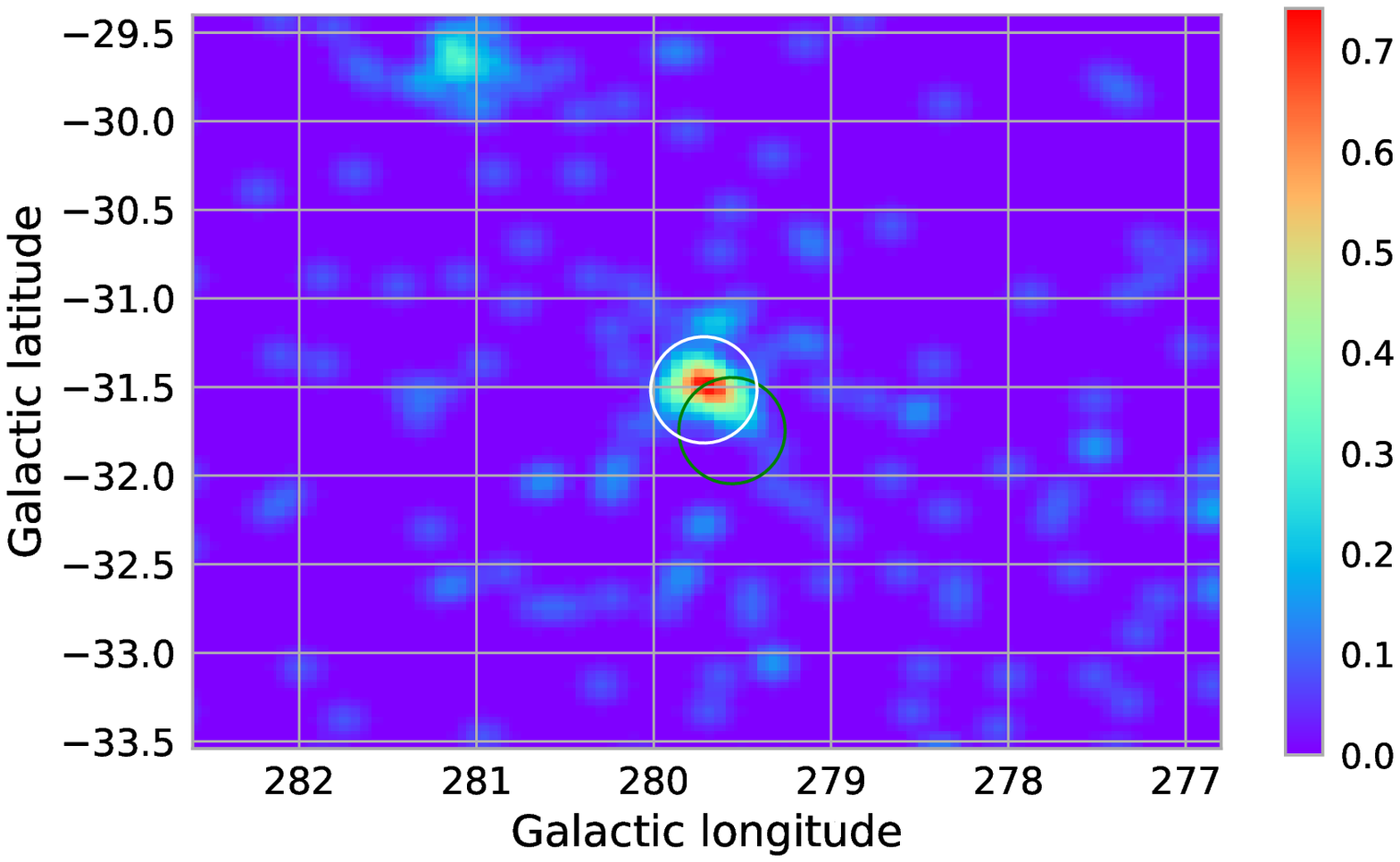}
	\includegraphics[width=\columnwidth]{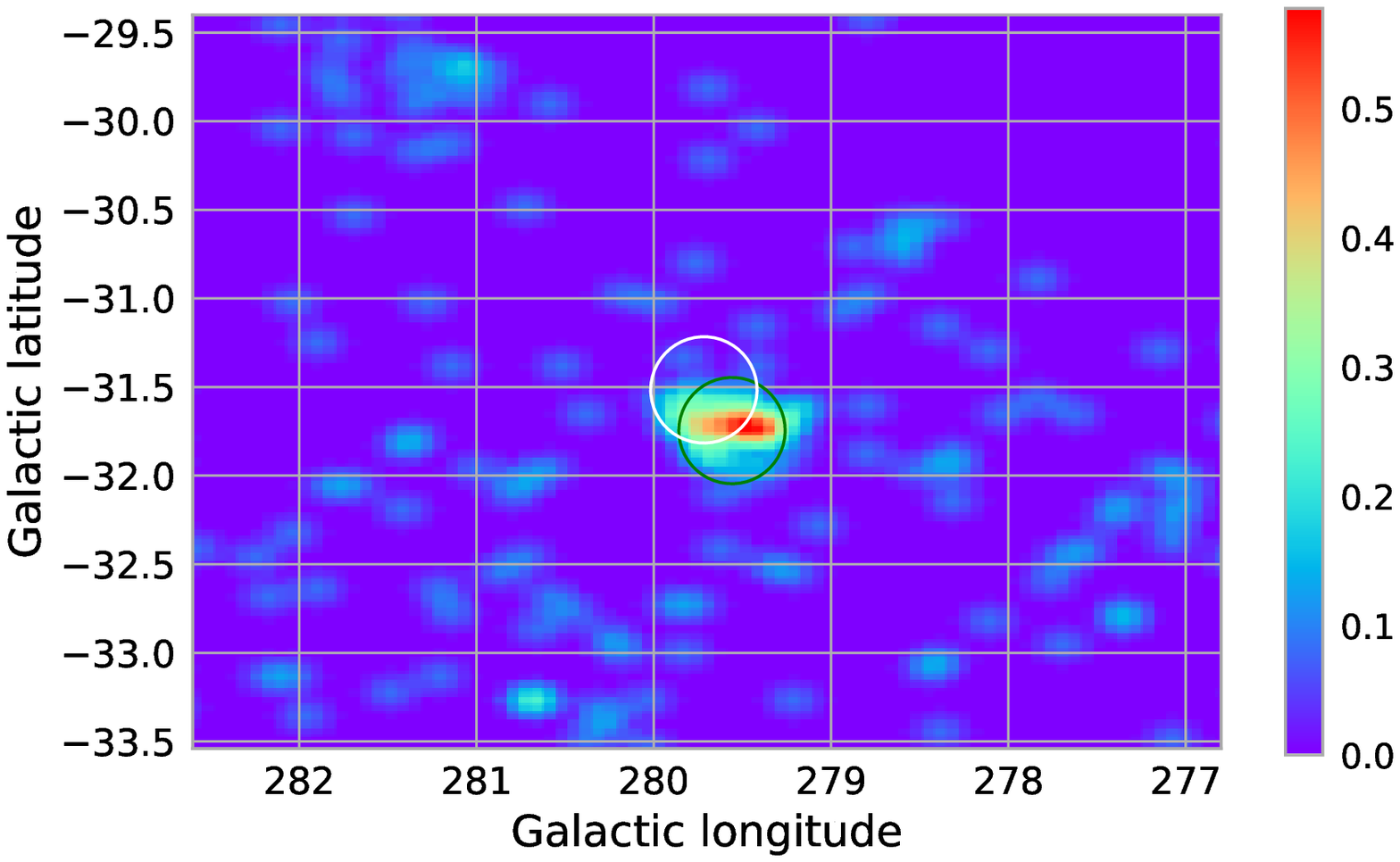}
    \caption{The count maps in 1--10~GeV. \textit{Upper panel.}  Eight weeks when PSR~J0540--6919 had $\mathrm{TS}>20$. \textit{Lower panel.} Four weeks when PSR~J0537--6910 had $\mathrm{TS}>20$. The circles are centered on the pulsars and are of $0.3\degree$ radius.}
    \label{fig:cmap1000}
\end{figure}

One can see from the plot of TS values for observations in 1-10 GeV energy range in Fig.~\ref{fig:ts_week} that the TS value of both PSR~J0540--6919 and PSR~J0537--6910 in some weeks can be as high as 20 while for the most of the time, they are barely detected. We found eight weeks when $\mathrm{TS}>20$ for PSR~J0540--6919, and four such weeks for PSR~J0537--6910. TS=20 threshold was chosen as a somewhat arbitrary compromise between significance  and number of detections.  Given that the number of trials (individual weeks) is so large we decided to set the threshold  rather high in order to select more robust detections. The weeks and the corresponding Fermi seconds are shown in Table~\ref{tab:ts1000}.
\begin{table}
	\centering
	\caption{Weeks with high TS ($>20$) of PSR~J0540--6919 and PSR~J0537--6910.}
	\label{tab:ts1000}
	\begin{tabular}{lccr} % four columns, alignment for each
		\hline
		 & Week & Start MET & End MET\\
		\hline
		\multirow{8}{*}{PSR~J0540--6919} & 65 & 278265118 & 278869918\\
		 & 88 & 292175518 & 292780318\\
		 & 236 & 381685918 & 382290718\\
		 & 257 & 394386718 & 394991518\\
		 & 548 & 570383518 & 570988318\\
		 & 590 & 595785118 & 596389918\\
		 & 649 & 631468318 & 632073118\\
		 & 662 & 639330718 & 639935518\\
		 \hline
		 \multirow{4}{*}{PSR~J0537--6910} & 542 & 566754718 & 567359518\\
		 & 587 &  593970718 & 594575518\\
		 & 601 &  602437918 & 603042718\\
		 & 624 &  616348318 & 616953118\\
		 \hline
	\end{tabular}
\end{table}
For these weeks, we constructed the count maps shown in Fig.~\ref{fig:cmap1000}.
One can see that, when one source is bright, the other one is not visible. In other words, at energies $E>1~$GeV band we probably see the flaring activity in both PWNe separately. In these weeks, the flux can be determined quite well. The flux of PSR~J0540--6919 in the selected eight weeks is $(8.8 \pm 1.4) \times 10^{-9}\text{cm}^{-2}\text{s}^{-1}$, while the average flux is $2.2 \times 10^{-9}\text{cm}^{-2}\text{s}^{-1}$. The flux of PSR~J0537--6910 in the selected four weeks is $(8.8 \pm 1.7) \times 10^{-9}\text{cm}^{-2}\text{s}^{-1}$, and the average flux is $1.5 \times 10^{-9}\text{cm}^{-2}\text{s}^{-1}$. Thus, we observe the flux enhancements of a factor 4 and 6. It is worth noting that the weeks of enhanced brightness do not coincide with those in other energy bands.

One should bear in mind that the enhancement in the weeks 639--642 is above the flux of
\textit{both} PSR~J0540--6919 and PSR~J0537--6910 and \textit{both} PWNe around them. If we could separate the
contribution of the actual source of the flare, its relative amplitude might be substantially larger. This is
very unlike the situation with the Crab nebula where we not only have a single supernova remnant, but also
can effectively eliminate the contribution of the pulsar by means of the gating procedure when all the photons corresponding to phases of  pulses are discarded.
For the pulsars in the LMC, one can do the same if the ephemeris of the pulsars were known. We found several
works where the relevant data for PSR~J0540--6919 are provided for various time intervals
\citep{Ferdman2015, Marshall2016, Kim2019}, but unfortunately we failed to construct any meaningful
phase diagram. This is not surprising: the papers cited used X-ray ephemeris which spanned the time interval of interest. \cite{Ferdman2015} used a long interval of 15.8 years, but it ends in December 2011, i.e. the overlap with our LAT data is only about 3 years. \cite{Marshall2016} and \cite{Kim2019} used data sets of 16 months and 3 years respectively. For our long time interval of 13.3 years relevant X-ray ephemeris are absent, hence to perform gating adequately was problematic even for PSR~J0540--6919. However, one could construct a timing solution for a portion of the whole observation time span. In particular, we are most interested in the weeks 639--642. First, we tried to do it solely with the LAT data, but did not succeed: for short time intervals, the photon statistic is apparently too low to detect the pulse profile, even if present; for longer intervals phase coherence is important, but hard to establish. Therefore, we attempted to time PSR~J0540--6919 using X-ray data from NICER \citep{Gendreau2016}. We used all the available data which comprise 74 observations from 2017 to 2022. We reduced them with \textit{heasoft-6.28} and \mbox{NICERDAS} version 10 by running \textit{nicerl2} with the default parameters and \textit{barycorr}. We measured the frequencies for the individual observations and found pulse times of arrival (ToAs). Unfortunately, we could not build a phase coherent solution, probably because of insufficient accuracy of frequency determination, but we estimated the frequency and its first derivative from an incoherent analysis, i.e. by fitting a straight line to the frequency vs. time dependency. Still, applying this to the LAT data within the weeks 639-642 did not result in pulse detection.
Moreover, one can see from Figure~6 of \cite{Mignani2019} that the pulses in gamma
rays are rather broad and low-amplitude in this pulsar, therefore the effect of gating,
even if successful, is not expected to be substantial.

As to PSR~J0537--6910, we found ephemeris for the time interval of interest \citep{Ho2022}, but again the gating procedure could not reveal any pulsation.

Still, we can try and make some rough estimates.
According to 4FGL-DR2, PSR~J0540--6919 is about seven times brighter than PSR~J0537--6910. Assume that all the flux from the composite source is produced by PSR~J0540--6919 and the PWN is as bright as the pulsar. In Section~\ref{sec:results} we showed that the flux of the flare in the 300--1000~MeV band is $3.0 \times 10^{-8}\text{cm}^{-2}\text{s}^{-1}$. Note that this is not the total flux of the flaring source, it is the flux \textit{above} the quiescent level. The quiescent flux of PSR~J0540--6919 is $1.24 \times 10^{-8} \text{cm}^{-2}\text{s}^{-1}$. If one half of it is due to the PWN, then its flux was enhanced by a factor of six in the flare. Or, if the nebula provides only $1/4$ of the total flux like in Crab, then its enhancement in the flare is of a factor 10. In the 100--300~MeV band, where the fluxes of the flare and PSR~J0540--6919 are $7.78 \times 10^{-8}$ and $7.97 \times 10^{-8} \text{cm}^{-2}\text{s}^{-1}$, analogous considerations show that the flux of the nebula grew by a factor of 3 to 5.

As an additional check, we constructed the light curves with various time binnings: 2, 3 and 4 weeks. The results are shown in Appendix~\ref{appendix}. Qualitatively, the results are the same: the flare is most evident in the 300--1000~MeV band, but also visible in the 100--300~MeV band. In the 1--10~GeV band, the flare is absent, we do not show this band in the plot. One can see that, depending on the binning, there can appear spikes at various bins, but only the 639--642 week enhancement (and the flare on week 15 reported by \cite{Fermi-flare}) persists not only with different binnings, but also in the two energy bands.

From our analysis we can roughly characterize the spectrum of the flare. In the 300--1000~MeV band, the flare source has a spectral index of $2.4 \pm 0.6$ which is quite close to the index of PSR~J0540--6919 (2.5). In the 100--300~MeV band, the spectral index of the flare source is $1.1 \pm 0.7$. Despite rather large error, the spectrum is considerably hard. This can be compared to the giant flare in the Crab nebula analyzed by \cite{Buehler2012}. They found the spectral index of the flare $\gamma_F=1.27 \pm 0.12$. They modeled the spectrum by a power law with an exponential cutoff with the cutoff energy $E_c \lesssim 500$~MeV so that it naturally softens at higher energies. This is similar to what we observe in the LMC.

%%%%%%%%%%%%%%%%%%%%%%%%%%%%%%%%%%%%%%
%%%%%%%%%%% CONCLUSIONS %%%%%%%%%%%%%%
%%%%%%%%%%%%%%%%%%%%%%%%%%%%%%%%%%%%%%
\section{Conclusions}\label{sec:conclusions}
In this work we performed a search for flares in gamma rays in the pulsar wind nebulae located in the LMC.
The search is motivated by the resemblance between the properties of the pulsars in the LMC and that in
the Crab nebula where vigorous flaring activity has been observed. We used data collected by Fermi-LAT
in more than 13 years to construct weekly light curves in several energy bands, from 100~MeV to 10~GeV.
In the 300--1000~MeV band, we found a threefold enhancement in flux in the weeks 639--642 (2020 October 26 -- November 23) that is likely to be a flare. A more thorough analysis of these four weeks showed that this enhancement is real at a significance level of $6\sigma$. This enhancement could be produced by a flare in one of the PWNe, which then increased its luminosity by up to an order of magnitude. If the flare originates from PSR~J0540--6919 then the total fluxes of the source during the flare in 100--300 and 300--1000~MeV bands are $1.58 \times 10^{-7}$ and $4.24 \times 10^{-8}  \text{cm}^{-2}\text{s}^{-1}$ respectively. This flare would be brighter than the brightest flare observed from the Crab pulsar if observed from 2 kpc -- the Earth-Crab distance. The spectrum of the main flare we analyzed is qualitatively similar to that of the giant flare in Crab in April 2011. 

We checked several factors which could lead to a confusion, but they seem not to be relevant. The nearby flaring
blazar PKS~0601-70 is detected as a separate source and its flux changes independently from that of the
PWNe in the LMC. The unknown source responsible for a flare on the week 15 is not likely to host the flare
on the weeks 639--642 as our analysis shows.

Our analysis reveals distinctive signs of flaring activity in the 1--10~GeV range when the flux of PSR~J0540--6919 and PSR~J0537--6910 reach $8.8 \times 10^{-9} \text{cm}^{-2}\text{s}^{-1}$. Interestingly, this activity is not correlated with that at lower energies. This is different from what is observed in Crab. Statistical analysis of \cite{Huang2021} shows that flares in Crab occur at lower energies, and the high-energy cutoff in their spectral fits is always below 1~GeV. Our observations of PSR~J0540--6919 and PSR~J0537--6910 suggest there can be flux enhancements at higher energies only, without corresponding counterparts at lower energies.
\section*{Acknowledgements}
The authors want to thank Dr.~N.~Porayko for her help with the pulsar timing analysis.
The work of the authors was supported by the Ministry of Science and Higher Education of Russian Federation
under the contract 075-15-2020-778 in the framework of the Large Scientific Projects program within the national
project "Science". The numerical part of the work was done at the computer cluster of the Theoretical Division of the Institute for Nuclear Research of the Russian Academy of Sciences.

%%%%%%%%%%%%%%%%%%%%%%%%%%%%%%%%%%%%%%%%%%%%%%%%%%
\section*{Data Availability}
The analysis is based on data and software provided by the Fermi Science Support Center (FSSC).
This research has made use of NASA’s Astrophysics Data System.



\bibliographystyle{mnras}
\input{output.bbl}

%\begin{thebibliography}{99}
%\end{thebibliography}

%%%%%%%%%%%%%%%%%%%%%%%%%%%%%%%%%%%%%%%%%%%%%%%%%%

%%%%%%%%%%%%%%%%% APPENDICES %%%%%%%%%%%%%%%%%%%%%

\appendix
\section{Supplementary plots} \label{appendix}

\begin{figure*}
    \includegraphics{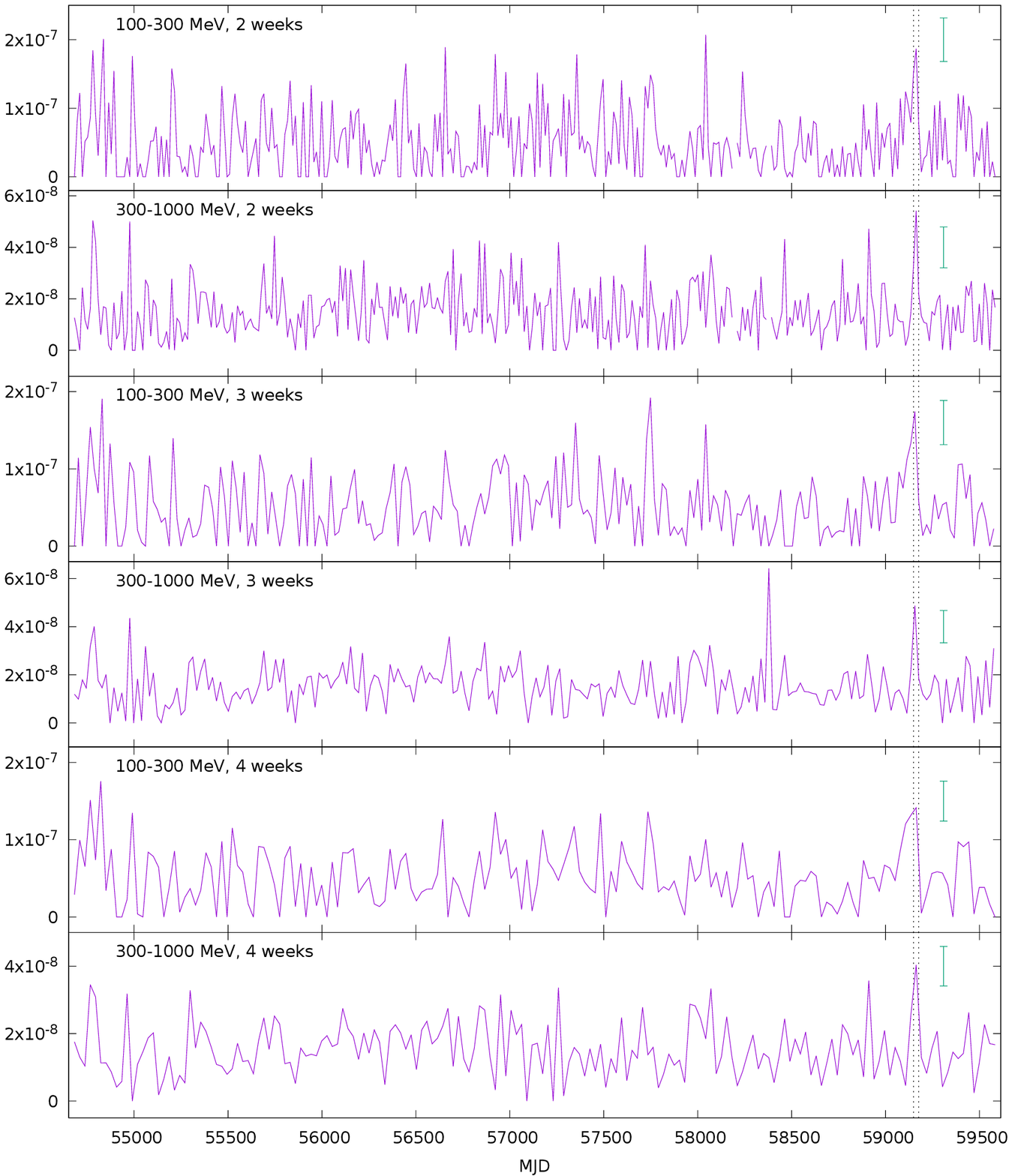}
    \caption{Light curves of the composite source in the bands 100--300 and 300--1000~MeV with different binnings indicated in the plots. The error bars indicate the median error in the corresponding plot. Vertical dashed lines indicate the weeks 639--642.}
    \label{fig:binnings}
\end{figure*}

In Fig.~\ref{fig:binnings} we show the light curves of the composite source in the bands 100--300 and 300--1000~MeV with different binnings.

%%%%%%%%%%%%%%%%%%%%%%%%%%%%%%%%%%%%%%%%%%%%%%%%%%

% Don't change these lines
\bsp	% typesetting comment
\label{lastpage}
\end{document}